\documentclass[journal=acs photonics,manuscript=article]{achemso}

\usepackage{amsmath}
\usepackage{geometry}%
\usepackage{amsfonts}%
\usepackage{amssymb}%
\usepackage{graphicx}

\usepackage{xcolor}

\author{S. Ali Hassani Gangaraj}
\affiliation[Unknown University]
{School of Electrical and Computer Engineering, University of wisconsin-Madison, Madison, Wisconsin 53706, USA}
\email{ali.gangaraj@gmail.com}

\author{Francesco Monticone}
\affiliation[Unknown University]
{School of Electrical and Computer Engineering, Cornell University, Ithaca, New York 14853, USA}
\email{francesco.monticone@cornell.edu}

\title[An \textsf{achemso} demo]
  {Drifting electrons: Nonreciprocal plasmonics and thermal photonics}

\begin{document}



\begin{abstract}
Light propagates symmetrically in opposite directions in most materials and structures. This fact -- a consequence of the Lorentz reciprocity principle -- has tremendous implications for science and technology across the electromagnetic spectrum. Here, we investigate an emerging approach to break reciprocity that does not rely on magneto-optical effects or spacetime modulations, but is instead based on biasing a plasmonic material with a direct electric current. Using a 3D Green function formalism and microscopic considerations, we elucidate the propagation properties of surface plasmon-polaritons (SPPs) supported by a generic nonreciprocal platform of this type, revealing some previously overlooked, anomalous, wave-propagation effects. We show that SPPs can propagate in the form of steerable, slow-light, unidirectional beams associated with inflexion points in the modal dispersion. We also clarify the impact of dissipation (due to collisions and Landau damping) on nonreciprocal effects and shed light on the connections between inflexion points, exceptional points at band-edges, and complex modal transitions in leaky-wave structures. We then apply these concepts to the important area of thermal photonics, and provide the first theoretical demonstration of drift-induced nonreciprocal near-field radiative heat transfer between two planar bodies. Our findings may open new opportunities toward the development of nonreciprocal magnet-free devices that combine the benefits of plasmonics and nonreciprocal photonics for wave-guiding and energy applications.
\end{abstract}


\section{Introduction}

The Lorentz reciprocity principle limits the performance of electromagnetic and photonic devices in a wide range of areas, from communication systems to quantum computing and thermal photonics. A reciprocal system, with its symmetric response when source and detector are interchanged, is not ideal when classical/quantum energy/information needs to be emitted, transferred, and absorbed unidirectionally with high efficiency, or when different information streams need to be selectively routed or processed based on their propagation direction. Standard nonreciprocal devices, such as magneto-optical isolators or circulators, are routinely used in many of these scenarios, but with the increasing demand for efficient electromagnetic and photonic integrated systems, there has been increasing interest in different solutions that permit asymmetric or, in the extreme case, truly unidirectional and beam-like wave propagation at small scales. In this context, particularly intriguing opportunities may be offered by combining the benefits of plasmonics (field localization and enhancement) with strong nonreciprocal effects \cite{FM_Nature}.


The oldest and most common approach to break reciprocity and achieve one-way light propagation is based on the magneto-optical effect, which requires biasing certain materials (plasmas, ferrites, magnetic garnets, etc.) with a static magnetic field \cite{gyro-1,gyro-2,gyro-3,gyro-4,gyro-5,Hassani_optica, PRL_Hassani_2, PRL_Hassani, Hassani_TAP, Hassani_PRApplied, Hassani_AWPL, Zeki,Shastri,Abdelrahman}. However, the need for a large external magnet, the weakness of the magneto-optical response at the relevant frequencies, and other practical issues, make this approach less appealing to realize strong nonreciprocal effects at the (sub)wavelength scale and for on-chip integration. Other approaches to break reciprocity have been extensively studied in the recent past, relying on nonlinear phenomena \cite{Nonlinear-1,Nonlinear-2,Nonlinear-3,Nonlinear-4,Nonlinear-5,Nonlinear-6,Nonlinear-7}, spatiotemporal modulations \cite{Spatio-1,Spatio-2,Spatio-3}, and multi-physics, e.g., optomechanical, interactions \cite{Optomech-1,Optomech-2,Optomech-4}. Unfortunately, however, the issues associated with dynamic reciprocity for nonlinear nonreciprocal devices, the typically narrowband response of optomechanical resonators, and the practical difficulties in implementing spatiotemporal modulations at high frequencies limit the applicability of these approaches. 

In addition to unidirectional wave-guiding, another particularly important area that may benefit from strong forms of nonreciprocity is thermal photonics \cite{Fan-Joule}. According to Kirchhoff's law of thermal radiation, the emissivity of a material is exactly equal to its absorptivity, at each frequency and angle. Rather remarkably, this law is not a consequence of the Second Law of Thermodynamics, but it originates directly from Lorentz reciprocity \cite{Landau-1,Fan-Joule,Axion} and is analogous to other relevant symmetries in electromagnetics, such as the fact that antennas made of reciprocal materials can be used in reception and transmission with exactly the same gain and directivity. One of the direct consequences of reciprocity in this context is that an efficient absorber will also be an efficient emitter. For instance, an absorber designed to efficently harvest solar radiation must radiate part of the energy back to the environment, which represents an intrinsic loss mechanism. This is the idea behind the Landsberg limit \cite{Landsberg}, which sets the upper efficiency bound for harvesting incoming solar radiation. 
This limit can only be reached with the use of nonreciprocal systems \cite{Landsberg,Ries,Green, Buddhiraju}. Moreover, breaking the symmetry between absorptivity and emissivity may also open new opportunities for near-field heat transfer. Nonreciprocal systems have been shown to enable thermal Hall effects \cite{Hall-effect} and the emergence of persistent heat currents in systems in thermal equilibrium at a certain temperature (analogous to persistent electrical or mass currents in superconductors or superfluids, respectively) \cite{Fan-S12_PRL}. Such nonreciprocal systems may become an important element in the quest for controlling heat flow at the nanoscale with large flexibility and efficiency. 
%
%
%
As in the case of nonreciprocal wave-guiding, nonreciprocal effects in thermal photonics are typically obtained by relying on magneto-optical materials \cite{Axion,Fan-S12,Ries,Miller-gyro,Zhu-Fan,Fan-Teslas}, which have been theoretically shown to enable a near-complete violation of Kirchhoff's law \cite{Zhu-Fan,Fan-Teslas} or nonreciprocal heat transfer between planar bodies \cite{Fan-S12}. Again, due to the weakness of magneto-optical effects at optical frequencies, these platforms require a relatively strong magnetic bias, which may raise questions about their practical applicability. A relevant exception is Ref. \cite{Y_Hadad}, where nonreciprocal emission/absorption was realized using space-time modulations; this strategy, however, was only demonstrated at microwave frequencies, in the form of a nonreciprocal leaky-wave antenna, and applying it to optical frequencies involves significant challenges.

In contrast with these previous studies, here we focus on a drastically different approach to break reciprocity in plasmonic platforms without the need for magneto-optical or dynamical effects. This approach is based on biasing certain conducting materials with a direct electric current, which, just like the magnetic field, is a quantity that is odd under time reversal and, therefore, it can be used to break reciprocity \cite{Landau-1}. This strategy can in principle be applied to either three-dimensional (3D) conducting materials (metals, degenerately doped semiconductors, and plasmas) or two-dimensional (2D) media such as graphene, and it has been the subject of increasing attention in the plasmonic literature \cite{Bliokh,Mario-ACS,Mario-Active,Mario-Negative-Landau,Mario-Nonlocal,Mario-Plasmonics,Collimated-SPP,Stauber,Levitov,Polini,Wenger}, with a very recent experimental demonstration at optical frequencies \cite{Basov}. Strong nonreciprocal effects emerge if the electron drift velocity $\boldsymbol{v}_d $ is large enough to affect the dispersion of surface plasmon-polaritons (SPPs), which is possible in certain high-mobility conducting materials (e.g., graphene). The effect of drifting electrons on SPP propagation can be qualitatively explained in an intuitive way: surface waves on plasmonic materials involve collective charge oscillations coupled to electromagnetic fields; these propagating waves are dragged or opposed by the drifting electrons, which implies that SPPs ``see'' different media when propagating along or against the current flow. In other words, it was argued in \cite{Bliokh} that, in a current-biased 3D plasmonic medium, the drifting electrons produce a Doppler shift in the isotropic material permittivity, i.e., $ \omega \rightarrow \omega - \boldsymbol{k} \cdot \boldsymbol{v}_d $, where $\boldsymbol{k}$ is the wavevector. Such a linear wavevector-dependence implies that the material response becomes nonlocal (spatially dispersive), as well as nonreciprocal since oppositely propagating waves see a different permittivity. In Section 2, we clarify that, even though this Doppler-shift argument is strictly valid only for longitudinal modes (in general, the permittivity becomes a spatially dispersive tensor \cite{Lifshitz}), under certain conditions it provides a good approximation of the exact results and allows for a drastic simplification of the analysis and simulations.


While all previous works on drift-induced nonreciprocity were mostly interested in the 1D problem of one-way SPP propagation along the current flow or on specific 2D materials such as graphene, here, in the first part of the paper, we use a 3D Green function formalism to elucidate the different regimes of nonreciprocal SPP propagation on the 2D surface of a generic 3D plasmonic platform, revealing a number of anomalous and extreme wave-propagation effects beyond unidirectional propagation. We also clarify how the unavoidable presence of losses in this plasmonic platform affects the nonreciprocal modal dispersion. Then, in the second part of this work, we investigate, for the first time to the best of our knowledge, the potential of current-biased nonreciprocal systems for thermal photonics, and we theoretically demonstrate a clear signature of drift-induced nonreciprocity in the canonical problem of near-field radiative heat-transfer between two planar bodies.

\section{Current-Induced Nonreciprocity in Three-Dimensional Plasmonic Platforms}

We consider the general problem of SPP propagation along a planar metallic-dielectric interface at $z = 0$, as shown in Fig. \ref{Fig1}(a). We employ a 3D Green function analysis (see Supplemental Material) to study both the dispersion properties of the supported SPP eigenmodes and their propagation properties in source-excited configurations, for both local (non-biased, reciprocal) and nonlocal (current-biased, nonreciprocal) plasmonic materials. We assume a laterally infinite interface between two different materials with permittivity $\epsilon _{1}$ for $z<0$ and $\epsilon_{2}$ for $z>0$ (permeability $\mu_1=\mu_2=\mu_0$), and source/observation point above the interface (Fig. \ref{Fig1}(a)). 
\begin{figure}[h!]
	\centering\includegraphics[width=\columnwidth]{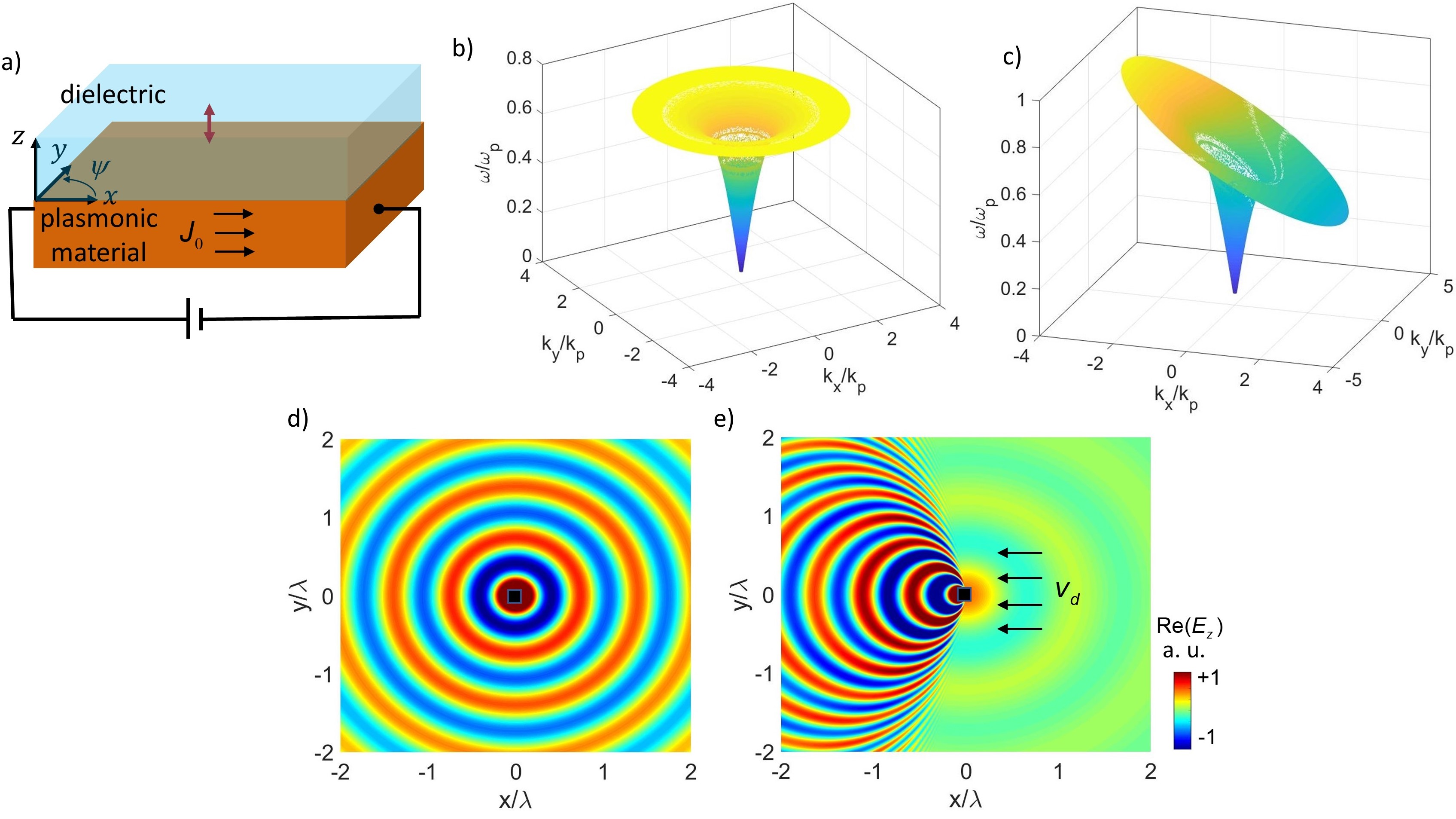}
	\caption{(a) Geometry under consideration: a plasmonic material is interfaced with a transparent dielectric. A vertically polarized point dipole source radiates close to the interface in the dielectric region. The supported SPP dispersion surface is shown in (b) for a reciprocal isotropic plasmonic material, with no current bias, interfaced with free space, and (c) for the same configuration but with the plasmonic material biased by a direct current with drift velocity $ \mathbf{v}_d = -c/10 \hat{x} $. (d,e) Source-excited in-plane distribution of the SPP electric field (time snapshot of the $z$ component) in the reciprocal and nonreciprocal cases: (d) without DC current at frequency $\omega/\omega_p= 0.6$, and (e) with DC current at $ \omega/\omega_p = 0.8 $. For panels (d) and (e) the source is located at $ z_0 = \lambda/10 $ above the interface, where $ \lambda $ is the free-space wavelength at the radiation frequency. The black square at the center denotes the source position. Animations of the calculated time-harmonic electric field distribution, for the reciprocal and nonreciprocal cases, are included as Supplemental Material.}\label{Fig1}
\end{figure}

For the transparent dielectric region we consider vacuum and for the metallic/plasmonic region, in the local non-biased case, we employ a standard Drude model for a free-electron gas with permittivity $ \epsilon(\omega) = 1 - (\omega_{p}/\omega)^2/(1+i\Gamma/\omega)$, where $ \omega_p $ is the plasma frequency and $\Gamma$ is the damping rate. In the following, we first study the problem with $\Gamma=0$ (lossless case), whereas the impact of dissipation is discussed in details in Section 2.3. At a local, isotropic, metal-vacuum interface, SPPs are known to exist for frequencies $ \omega \leq \omega_{SPP} = \omega_p/\sqrt{2} $. The 3D dispersion surface of these SPP eigenmodes, calculated by tracking the poles of the Green function integrand in Eq. (33) of the Supplemental Material, is shown in Fig. \ref{Fig1}(b) for the lossless case. The dispersion surface clearly exhibits a flat dispersion asymptote at the surface plasmon resonance $ \omega = \omega_{SPP} $, as expected for a local/lossless Drude model \cite{Novotny} (see, e.g., \cite{FM_Nature,Hassani_optica} and references therein, for a discussion of how losses and nonlocal effects modify this dispersion diagram). The rotationally symmetric shape of the dispersion surface with respect to in-plane wavevectors indicates that the surface mode propagates reciprocally and isotropically on the interface. The spatial field distribution of the SPPs excited by a vertically polarized source as in Fig. \ref{Fig1}(a) was calculated using our 3D Green function formalism at $ \omega=0.6\omega_p $ (see Fig. \ref{Fig1}(d)), confirming that surface waves propagate symmetrically and omnidirectionally along the interface.

As discussed in Ref. \cite{FM_Nature}, strong nonreciprocity and unidirectionality in a continuous plasmonic platform of this type can be achieved if the mechanism that breaks reciprocity introduces an asymmetry in the dispersion asymptote for large wavevectors. While this is typically achieved with a static magnetic bias, a direct current bias, which can be directly carried by plasmonic materials due to their conductive nature, can also introduce a strong asymmetry, but with qualitatively different characteristics, as originally discussed in \cite{Bliokh}.The presence of a DC current means that conduction electrons move with average drift velocity $ \boldsymbol{v}_d = \boldsymbol{J}_0/ne $ where $\boldsymbol{J}_0$ is the current density, $ n = m\omega_p^2/4\pi e^2 $ is the electron density, and $ e $ is the negative electron charge. Ref. \cite{Bliokh} argues that the movement of electrons produces a Doppler frequency shift in the material permittivity, i.e., $ \epsilon(\omega - \boldsymbol{k} \cdot \boldsymbol{v}_d ) = 1 - \omega_p^2 / (\omega - \boldsymbol{k} \cdot \boldsymbol{v}_d)^2  $, such that the material response becomes spatially dispersive, with a wavevector-dependent isotropic permittivity. This point, and especially the isotropy assumption, is worth some clarification. As discussed in \cite{Lifshitz,Bittencourt}, in the presence of any non-trivial electron velocity distribution (due to a current bias or just the thermal motion of free electrons), the permittivity of a free-electron gas becomes spatially dispersive (in other words, the material response is nonlocal due to the electrons moving a certain distance during one period of the applied electromagnetic field). Importantly, in the presence of spatial dispersion, the material permittivity is a tensor even in an isotropic medium. The general form of this wavevector-dependent permittivity tensor can be written (in tensor notation in an arbitrary coordinate system) as \cite{Lifshitz},


\begin{equation}
	\epsilon_{ij}(\omega, \boldsymbol{k} ) = \epsilon_T(\omega, \boldsymbol{k} ) \delta_{ij} + \left(\epsilon_L(\omega, \boldsymbol{k}) - \epsilon_T(\omega, \boldsymbol{k})  \right)  k_ik_j/k^2
\end{equation}
where the scalar functions $ \epsilon_T $ and $ \epsilon_L $ are the transverse and longitudinal permittivities (for electric fields orthogonal or parallel to $\boldsymbol{k}$, respectively), $k$ is the wavevector magnitude, $k_{i}k_{j}$ denotes a tensor product between the wavevector and itself, and $\delta_{ij}$ indicates the identity matrix. In Supplemental Material (Notes I and II), we derive $ \epsilon_T $ and $ \epsilon_L $ starting from the Vlasov equation for plasmas, and we show that only the longitudinal permittivity is perfectly Doppler-shifted in the presence of a drift current, whereas the transverse permittivity exhibits a different, weaker form of spatial dispersion. Since the electric field of a SPP mode on a 3D plasmonic material has both transverse and longitudinal components inside the material, its dispersion and propagation properties are expected to depend on both $ \epsilon_T $ and $ \epsilon_L $. (in a 2D electron gas, instead, it is clear that only the longitudinal conductivity is relevant even if the field has a transverse out-of-plane component). This implies that the modeling approach proposed in \cite{Bliokh}, which uses the same equation for SPP dispersion as in the local unbiased case, but substituting the longitudinal Doppler-shifted permittivity, can only be approximately valid. To clarify this point, in the Supplemental Material (Note I), we derive the exact modal solution for SPPs propagating along the electron current and compare it against the SPP dispersion obtained when only the longitudinal or transverse permittivity is considered. Our findings show that, while one has to be aware of this issue for more accurate calculations, the model proposed in \cite{Bliokh} using only the longitudinal permittivity is approximately correct and captures the relevant physics especially in the large-wavevector regime. Based on these considerations, we can therefore safely use this approximate model to drastically simplify the analysis and numerical calculations, and provide additional physical insight. Furthermore, it should be mentioned that, while the nonlocal model based on the Doppler effect is qualitatively valid for the 3D plasmonic materials considered here, its validity for 2D systems, such as graphene, has been the subject of debate in the literature \cite{comment-on}, and alternative models have been proposed in \cite{Stauber,Levitov,Polini,Wenger,Svintsov}. However, it has been argued that, even for 2D systems, this model is accurate when the electron-electron collisions predominate and force the electron gas to move with approximately constant velocity, which makes the drift-biased medium act similarly to a moving medium \cite{Mario-Active,Reply}. 


While Ref. \cite{Bliokh} focused on 1D propagation in the direction of the current, here we use our 3D Green function formalism to investigate drift-induced nonreciprocal effects over the entire two-dimensional interface of the three-dimensional geometry in Fig. \ref{Fig1}(a), revealing anomalous and extreme forms of surface-wave propagation on this surface. The modified dispersion and propagation properties of the SPPs supported by this nonreciprocal configuration can be readily analyzed using the same 3D Green function as for the reciprocal case, but, as explained above, with the nonlocal, Doppler shifted, longitudinal permittivity, $ \epsilon_L (\omega , \boldsymbol{k}) = 1 - \omega_p^2 / (\omega - \boldsymbol{k} \cdot \boldsymbol{v}_d)^2  $. As a first example, we suppose the electron flow is along the $+x$-axis, with drift velocity $v_d = -c_0/10$ (where $c_0$ is the speed of light in vacuum). We must emphasize that this value of the drift velocity is unrealistically high and is considered here just for illustrative purposes to describe the qualitative behavior of surface waves in this nonreciprocal platform. Later in the article, and in the Supplemental Material, we provide results for much smaller values of drift velocity. The modified SPP dispersion surface is shown in Fig. \ref{Fig1}(c). The dispersion asymptote is now tilted as a result of the Doppler shift, as discussed in \cite{Bliokh}, making the dispersion surface strongly asymmetric and nonreciprocal in the $x$-direction, with a frequency range where SPPs can only propagate toward the negative $x$-axis. In addition, unidirectional SPPs are now supported at frequencies where SPPs could not propagate at all in the reciprocal case (frequencies higher than the flat asymptote in Fig. \ref{Fig1}(b)). The field distribution of surface waves launched by a vertical dipole source is shown in Fig. \ref{Fig1}(e) for a frequency higher than the original surface-plasmon resonance, $ \omega = 0.8 \omega_p > \omega_{SPP} $. We clearly see that, in this case, SPPs propagate preferentially toward left, parallel to the electrons drift velocity, despite the homogeneity and isotropy of the surface. While this behavior is somewhat expected as an extension of the 1D results of Ref. \cite{Bliokh} to a 2D surface, even more extreme wave-propagation effects emerge at different frequencies and angles, as discussed in the next sections.  

\subsection{Inflexion Points and Slow-Light Beaming}

\begin{figure}[h!]
	\centering\includegraphics[width=\columnwidth]{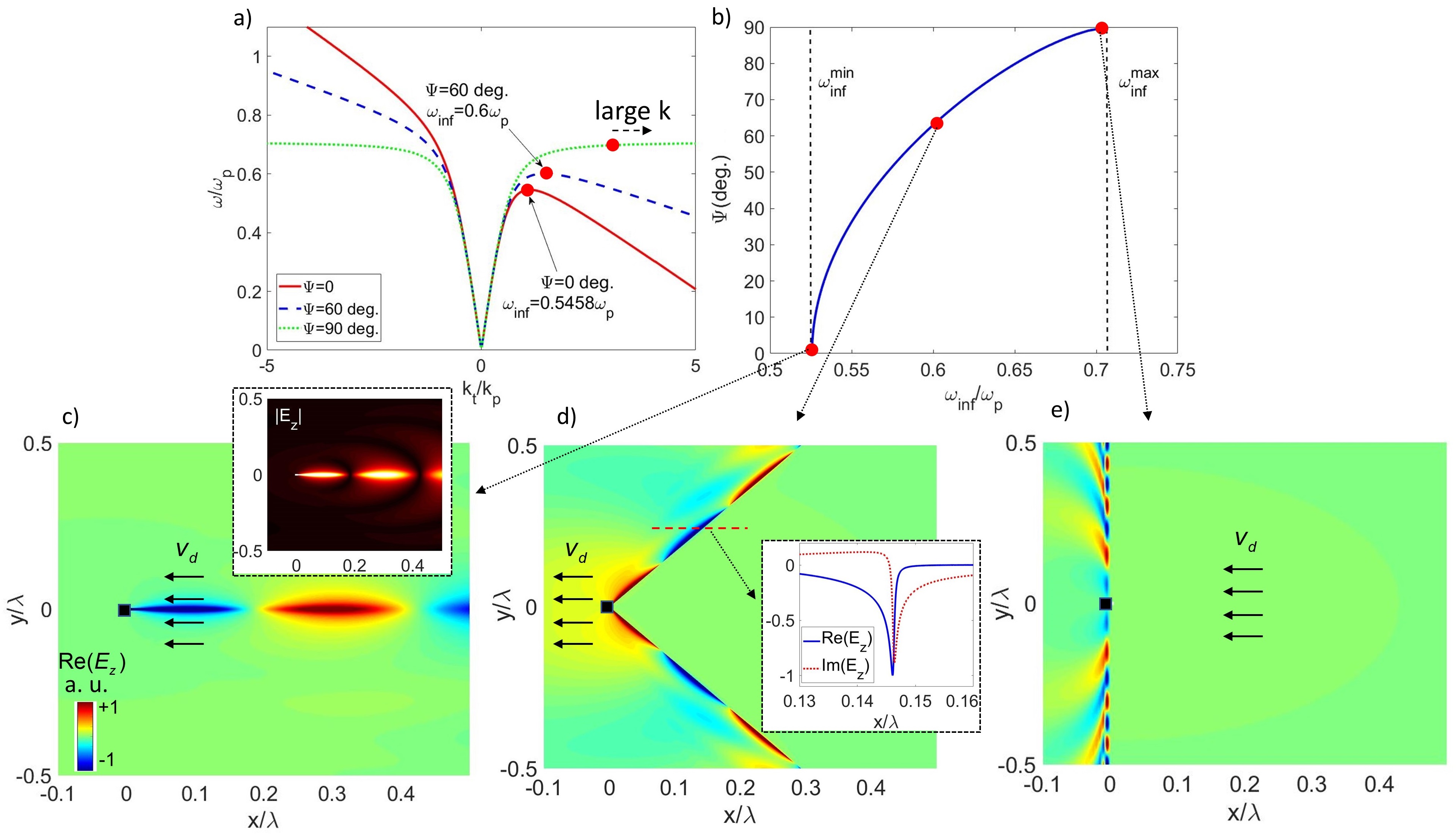}
	\caption{(a) Dispersion curves for SPPs propagating along different angles with respect to the current flow. The angle $ \Psi $ is measured from the positive $x$-axis (see Fig. \ref{Fig1}(a)). Different curves correspond to the intersection of the dispersion surface in Fig. \ref{Fig1}(c) with planes at angle $ \Psi $ with respect to the $ k_x $-axis. The red dots indicate inflexion points for different directions of propagation. (b) Inflexion-point frequencies for SPPs propagating along different angles. (c,d,e) Source-excited in-plane distribution of the SPP electric field (time snapshot of the $z$ component) for (c) $ \omega = \omega_{inf}^{min} $, (d) $ \omega = 0.6\omega_p $ and (e) $ \omega \approx \omega_{SPP} = \omega_{inf}^{max} $. The $x$ and $y$ axes are normalized with respect to the free-space wavelength at the radiation frequency. The black squares indicate the in-plane position of the source, which is located in the vacuum region at $ z_0 = \lambda/10 $ above the interface for panels (c,d) and at $ z_0 = \lambda/30 $ for panel (e) (the source is closer to the interface in this panel to efficiently excite the high-wavevector components of the SPPs). The inset in panel (c) shows the amplitude distribution of the $z$-component of the electric field, and the inset in panel (d) shows the real and imaginary parts of $E_z$ along the red dashed line crossing the narrow beam in the main panel.}\label{Fig2}
\end{figure}

The results in Fig. \ref{Fig1} clearly show that the dispersion properties of drift-biased SPPs strongly depend on the angle $ \Psi $ between in-plane wavevector and current flow, as expected from the spatially dispersive form of the permittivity. To clarify this point further, Figure \ref{Fig2}(a) shows 1D dispersion curves for SPPs propagating along different angles. SPPs propagating normal to the current flow ($ \Psi = 90 $ deg.) do not interact with the drifting electrons and the dispersion curve is the same as in the reciprocal case with flat asymptotes at $ \omega = \omega_{SPP} $. Conversely, SPPs propagating along the current flow ($ \Psi = 0 $) are maximally affected by the presence of the current, particularly in the large-wavevector asymptotic region where spatial dispersion becomes particularly strong. 
The tilted dispersion asymptote is associated with highly localized SPPs dragged by the flow of electrons \cite{Bliokh}, and its slope (group velocity) is indeed exactly equal to the electron drift velocity seen by the SPP mode, i.e., $ \partial \omega / \partial k = \boldsymbol{v}_d \cdot \hat{u}_k $, where $ \hat{u}_k $ is the in-plane unit wavevector. Such a tilted asymptote makes the dispersion curve non-monotonic for wavevectors opposite to $\boldsymbol{v}_d$, with an \emph{inflexion point} (red dot in Figure \ref{Fig2}(a)) where the SPP group velocity vanishes and changes sign. The position of the inflexion point along the frequency axis depends on the direction of propagation, such that, for each angle $ \Psi$, an inflexion point exists at a different frequency,


\begin{align}
	\omega_{inf}& = \frac{\omega_p}{\sqrt{2}}\left[1 - \frac{3}{2}\left( \frac{v_d \cos(\Psi)}{2c} \right)^{2/3}\right] \nonumber \\& \Rightarrow 
	\begin{cases}
		\omega_{inf}^{min} =\frac{\omega_p}{\sqrt{2}}\left[1 - \frac{3}{2}\left( \frac{v_d}{2c} \right)^{2/3}\right], & \Psi = 0\\
		\omega_{inf}^{max} \rightarrow \frac{\omega_p}{\sqrt{2}},              & \Psi \rightarrow 90~ \mathrm{deg.} 
	\end{cases}
\end{align}
Thus, while an inflexion point exists for any angle $ \Psi \in [0,90)$ deg., its frequency is limited within the range $ \omega_{inf}^{min} \leq \omega_{inf} \leq \omega_{inf}^{max}  $, as shown in Fig. \ref{Fig2}(b). Minimum inflexion-point frequency occurs for SPP propagation along the current flow and maximum frequency is asymptotically reached for propagation normal to the current, in which case the inflexion point rapidly migrates to large values of wavevector as seen in Fig. \ref{Fig2}(a). For any frequency outside this window, no inflexion point appears, i.e., the SPP group velocity does not vanish along any direction. 

The existence of an angle-dependent inflexion point in this frequency window leads to nontrivial propagation effects on the 2D surface of the considered nonreciprocal platform. These effects, which have not been studied previously, can be elucidated by considering the system's Green function under quasi-static approximation.
Indeed, while the 3D Green function described above provides a complete formulation to study SPP propagation, it is usually evaluated numerically and tends to hide physical insight. A simpler, more revealing, approximate formulation can be obtained by neglecting retardation effects (quasi-static approximation) and assuming that the main radiation channel of the dipolar source is represented by the excitation of a single, guided, surface mode. As was shown in Ref. \cite{Diffractionless}, under these approximations, the source radiation intensity in a certain in-plane direction (radiated power per unit angle) is given by

\begin{equation} \label{rad_patt}
	U(\Psi) \approx \frac{\omega^2}{16 \pi} \frac{1}{|\boldsymbol{\nabla}_{\boldsymbol{\mathrm{k}}_t} \omega(\textbf{k}_t)|} \frac{1}{C(\textbf{k}_t)} | \boldsymbol{\gamma}^* \cdot \textbf{E}_{\textbf{k}} (z_0) |^2,
\end{equation}
where $ \textbf{k}_t $ is the in-plane wavevector, $ \tan (\Psi) = \mathrm{k}_y/\mathrm{k}_x $, $\textbf{E}_{\textbf{k}} (z_0)$ is the modal electric field at the location $z_0$ of a source with polarization state $ \boldsymbol{\gamma} $, $\omega(\textbf{k}_t)$ is the dispersion relation of the relevant mode (hence, $ \boldsymbol{\nabla}_{\boldsymbol{\mathrm{k}}_t} \omega(\textbf{k}_t)$ is the group velocity), and $C(\textbf{k}_t) $ is the local curvature of the equifrequency contour (EFC) of the dispersion relation. Equation (\ref{rad_patt}) gives the approximate in-plane radiation pattern of the dipole source, corresponding to the in-plane pattern of SPP propagation. This equation reveals that the SPP pattern can be controlled in two ways: (i) by suitably selecting the polarization of the dipole source, which controls the coupling factor $ | \boldsymbol{\gamma}^* \cdot \textbf{E}_{\textbf{k}} (z_0) |^2 $ between the source and the surface mode of interest; or (ii) by engineering the dispersion relation of the relevant surface mode, namely, by controlling the local curvature $ C(\textbf{k}_t) $ of the equifrequency contour and/or the wavevector dependence (angular dependence) of the group velocity, $ |\boldsymbol{\nabla}_{\boldsymbol{\mathrm{k}}_t} \omega(\textbf{k}_t)| $. In the considered current-biased plasmonic platform, the presence of inflexion points where the group velocity vanishes is therefore a crucial factor affecting the in-plane propagation pattern. Indeed, at any frequency where an inflexion point appears, the inverse of the group velocity in Eq. (\ref{rad_patt}) diverges, $1/ | \boldsymbol{\nabla}_{\boldsymbol{\mathrm{k}}_t} \omega(\textbf{k}_t)| \rightarrow \infty $ (in practice it becomes large but finite) along a specific angle with respect to the current flow. Thus, the in-plane propagation pattern is maximized and localized around this angle, resulting in a peculiar ``beaming effect'' as further discussed below.

Figure \ref{Fig2}(c,d,e) show the in-plane field distributions, calculated using our Green function formulation, for SPP propagation at three frequencies: (c) minimum and (e) maximum inflexion-point frequencies and (d) another frequency in between. As shown in Fig. \ref{Fig2}(a,b), minimum inflexion-point frequency corresponds to vanishing SPP group velocity in the direction of the current flow. This leads to a diverging behavior in Eq. (\ref{rad_patt}) at $ \Psi = 0 $, which implies that the SPP fields form a unidirectional narrow beam along the current flow, as confirmed by Fig. \ref{Fig2}(c). Note that no energy actually propagates toward the $+x$-axis since the group velocity is zero (group velocity and energy velocity are virtually identical in a low-loss medium even if material dispersion is high), and the narrow beam resembles a standing wave with localized and enhanced fields, as seen in the field amplitude distribution plotted in the inset of Fig. \ref{Fig2}(c) (an animation of the calculated time-harmonic electric field, which further clarifies this peculiar behavior, is included as Supplemental Material). Choosing a different frequency, for example $ \omega = 0.6 \omega_p $, an inflexion point appears for SPPs propagating at an angle $\Psi = 60$ deg. with respect to the current flow (see Fig. \ref{Fig2}(a), dashed blue line). This leads to narrow SPP beams along $ \Psi= \pm 60 $ deg., as shown in Fig. \ref{Fig2}(d). In this scenario, two beams emerge because the system's response is reciprocal normal to the current, therefore SPPs propagate symmetrically along the $y$-axis. As in the previous case, these beams are narrow standing waves with enhanced and localized fields. Interestingly, in this case the field distribution is even sharper around $ \Psi= \pm 60 $ deg., especially toward smaller angles, where no surface-wave propagating solution exists at this frequency (see Fig. \ref{Fig1}(c) and the equifrequency-contour animation in Supplemental Material). We note that, although the field pattern looks discontinuous around this angle, a closer inspection reveals that, rather than an actual field discontinuity (which would be nonphysical on a homogeneous surface), the fields decay very sharply, within a small fraction of a wavelength, as demonstrated in the inset of Fig. \ref{Fig2}(d) which shows the real and imaginary parts of the electric field along the red dashed line. We speculate that such ultra-sharp fields on a homogeneous surface may be useful for surface-sensing applications or optical trapping near the surface. Finally, Fig. \ref{Fig2}(e) shows the fields at $ \omega \approx \omega_{SPP} = \omega_{inf}^{max} $, at which an inflexion point appears for SPPs propagating close to $ \Psi= \pm 90 $ deg. Indeed, in this case, the field distribution is dominated by the presence of narrow beams almost normal to the current flow and with large wavenumber, consistent with Fig. \ref{Fig2}(a). 

The emergence of these narrow, current/frequency-steerable, frozen-light beams with enhanced fields is highly nontrivial, especially given the fact that the plasmonic platform that supports them is homogeneous, and they may offer new opportunities for tunable, directional, enhanced, light-matter interactions on plasmonic surfaces.

\subsection{Unidirectional Propagation and Negative Group Velocity}
\begin{figure}[h!]
	\begin{center}
		\noindent
		\includegraphics[width=0.8\columnwidth]{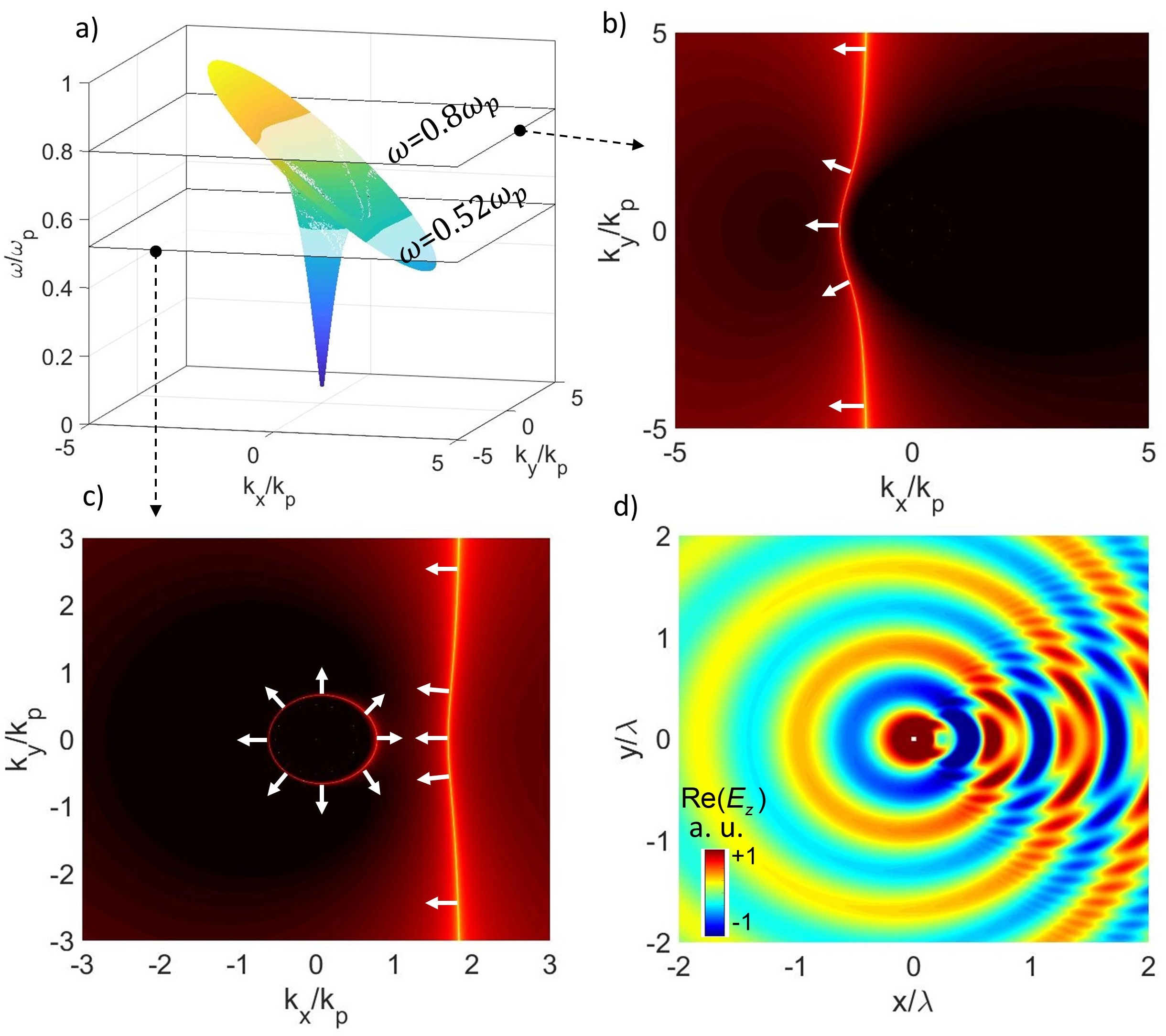}
		\caption{(a) SPP dispersion surface for $ v_d = -c/10 $ with two cut-planes at constant frequency, $ \omega = 0.8 \omega_{p} $ and $ \omega = 0.52\omega_p $. The cross sections correspond to the equifrequency contours (EFC) shown in panels (b) and (c). EFCs are represented as density plots of the magnitude of the integrand of the 3D Green function in Eq. (33) of the Supplemental Material. The source-excited electric field distribution (time snapshot of the $z$ component) at $ \omega = 0.8 \omega_{p} $, corresponding to panel (b), is shown in Fig. \ref{Fig1}(e), whereas panel (d) shows the field distribution at $ \omega = 0.52\omega_p $, corresponding to the EFC in panel (c).}\label{Fig3}
	\end{center}
\end{figure}

As discussed above, for frequencies higher than $ \omega_{inf}^{max} = \omega_{SPP} $ or lower than $ \omega_{inf}^{min} $, no inflexion point appears, and hence the SPP group velocity never vanishes along any in-plane direction. In these frequency regions, it is mostly the curvature of the equifrequency contour $ C(\textbf{k}_t) $ that determines, according to Eq. (\ref{rad_patt}), how SPPs propagate on the two-dimensional interface. Other anomalous wave-propagation effects occur in this regime, qualitatively different from the behavior of SPPs in the $ \omega_{inf}^{min} < \omega < \omega_{inf}^{min}$ window. Figure \ref{Fig3}(a) shows the dispersion surface with two cut-planes at constant frequencies, $ \omega = 0.8 \omega_{p} > \omega_{SPP} $ and $ \omega = 0.52\omega_{p} < \omega_{inf}^{min} $. The intersection of these planes with the dispersion surface gives the EFCs presented in Fig. \ref{Fig3}(b) and \ref{Fig3}(c), respectively (an animation of how the EFC changes as a function of frequency is included as Supplemental Material). The colors of the density plots correspond to the magnitude of the integrand in Eq. (33) of the Supplemental Material (brighter colors mean higher intensity), indicating which portions of the EFC contribute more strongly to SPP excitation for the chosen source, i.e., a dipolar emitter linearly-polarized along the $z$-axis and located above the interface as detailed in the caption of Fig. \ref{Fig1}. The white arrows, normal to the EFC at each point, indicate the main directions of energy flow (group velocity vectors). As Fig. \ref{Fig3}(b) suggests, at $ \omega = 0.8 \omega_{p}$ SPPs are expected to propagate unidirectionally toward the $-x$-axis (propagation along the $+x$-axis is prohibited) with wavefront spreading (diffraction) along the $y$-axis. The field distribution in Figure \ref{Fig1}(e) confirms this behavior. Further comments on this propagation pattern and on the impact of the source height from the surface are discussed in the Supplemental Material. 

The equifrequency contour has a more complicated non-connected shape for frequencies lower than $ \omega_{inf}^{min} $ due to the tilted dispersion asymptote. As seen in Fig. \ref{Fig3}(c), the contour is formed by a central ellipse slightly stretched toward right (in the absence of current bias, the central ellipse becomes a circle) and an approximately vertical line at larger positive values of $k_x$. This vertical line corresponds to the large-wavevector SPPs dragged backward by the drifting electrons, propagating with group velocity almost exactly equal to the drift velocity, as discussed in the previous sections. If the drift velocity is reduced, the vertical line moves to even larger values of $k_x$, going asymptotically to infinity for $v_d \rightarrow 0$. As the direction of the white arrows in Fig. \ref{Fig3}(c) suggests, small-wavevector SPPs are allowed to propagate along every in-plane direction with a slight preference toward the $ +x $-axis (ellipse contribution), but the dipole radiation may also couple to counter-propagating SPPs with positive $k_x$ but negative group velocity (vertical line contribution). Figure \ref{Fig3}(d) shows the corresponding source-excited field distribution at $ \omega = 0.52\omega_p $, demonstrating an asymmetric SPP wavefront, stronger toward the positive $ x $-axis, and characterized by a clear interference pattern due to the two contributions in Fig. \ref{Fig3}(c). An animation of the calculated time-harmonic electric field, corresponding to Fig. \ref{Fig3}(d), is included as Supplemental Material. At lower frequencies, the EFC and field distribution are qualitatively similar to those in Fig. \ref{Fig3}(c,d), but SPP propagation tends to become more symmetric and isotropic as frequency is lowered, consistent with the fact that the elliptical EFC becomes smaller and rounder (smaller wavevectors imply weaker spatial dispersion), whereas the vertical-line contribution tails off as it moves to larger values of $k_x$ (hence, such large-wavevector SPPs become increasingly difficult to excite).

We also note that, while the results in this section and the previous section have assumed a unrealistically high value of drift velocity for illustrative purposes, the nonreciprocal effects demonstrated here can be achieved at much lower drift velocities (examples are provided in Section 3 and in the Supplemental Material). In principle, as long as the drift velocity is nonzero, the plasmonic material remains nonreciprocal, but the nonreciprocal behavior may become so weak as to be overshadowed by dissipative and other effects. The non-trivial impact of dissipation on the modal dispersion is discussed in the next section. Furthermore, other forms of nonlocality, e.g., of hydrodynamic origin, may have an impact if dissipation is sufficiently weak. While a complete analysis of these nonlocal effects will be the subject of a future work, we expect that similar tradeoffs may exist as in nonreciprocal magnetic-biased plasmonic systems \cite{FM_Nature, Hassani_optica, PRL_Hassani_2,Buddhiraju2020}.

\subsection{Modal Transitions, Exceptional Points, and Impact of Dissipation}

\begin{figure}[h!]
	\centering\includegraphics[width=0.9\columnwidth]{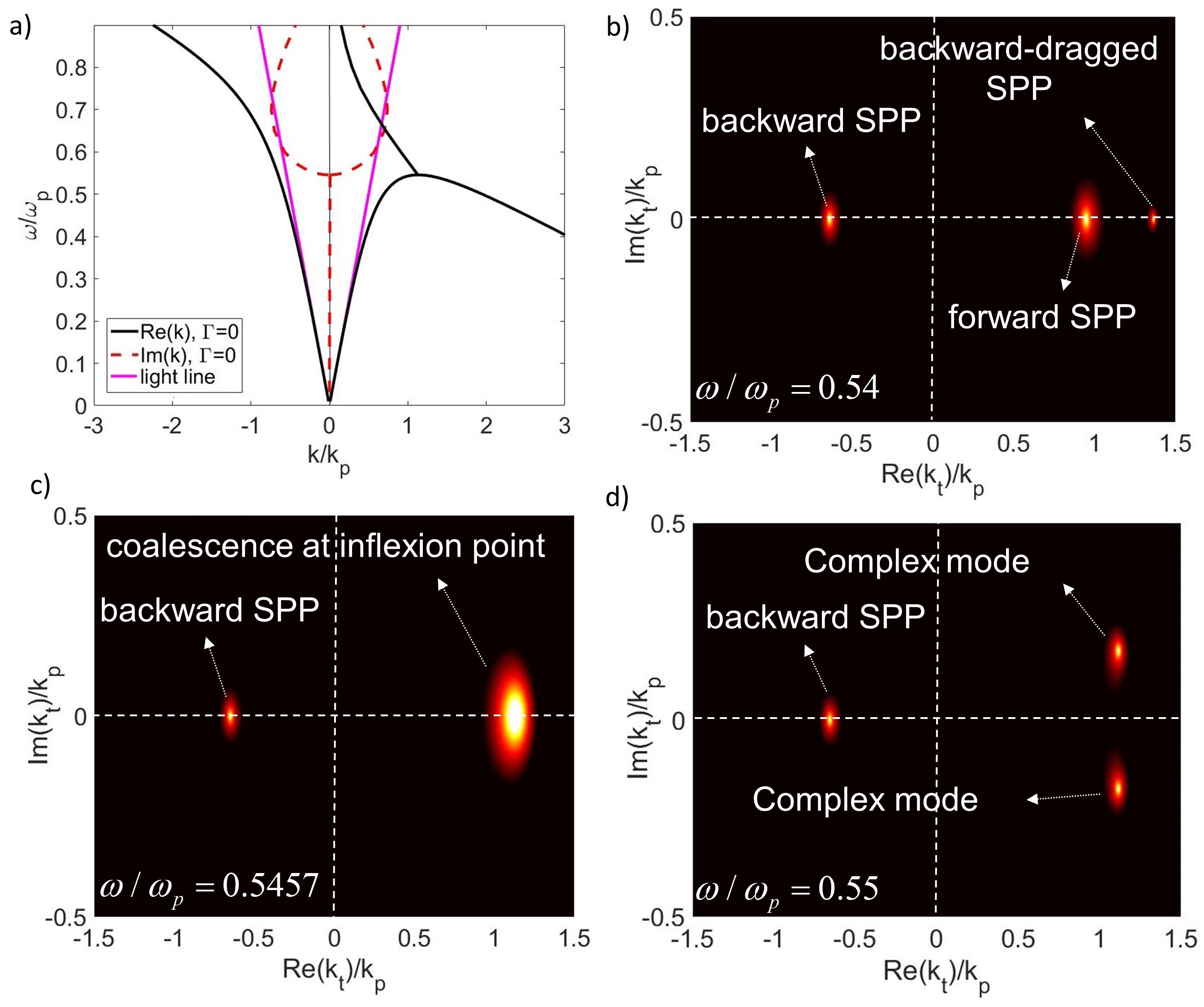}
	\caption{(a) Dispersion diagram for SPPs propagating in the direction of the current flow ($\Psi=0$), similar to Fig. \ref{Fig2}(a), but tracking each individual solution of the dispersion equation, and plotting its complex in-plane modal wavenumber, as frequency is varied. Absorption losses are assumed to be zero, i.e., $\Gamma=0$. See Fig. \ref{lossy} for the dissipative case. Solid purple lines indicate the light cone. (b,c,d) SPP pole constellation, on the complex $k_x$-plane (for fixed $k_y=0$), at different frequencies: (a) $ \omega/\omega_{p} = 0.54 < \omega_{inf}^{min} $, (b) $ \omega/\omega_{p} = 0.5457 = \omega_{inf}^{min} $ and (c) $ \omega/\omega_{p} = 0.55 > \omega_{inf}^{min} $, for SPP modes propagating along the drifting electrons with $ v_d = -c/10 $. The peculiar character of the two complex modes in panel (d) is discussed in the text.   
	}\label{EPDs}
\end{figure}

Since the inflexion point described above represents the most important feature of the dispersion diagram of a current-biased plasmonic system, marking the transition between different wave-propagation regimes, we further investigated the local dispersion behavior around this point and the impact of dissipation. 
Let us consider SPPs with propagation direction fixed parallel to the drift current, $\Psi = 0$, i.e., with their in-plane wavevector purely in the $x$-direction. The dispersion diagram for this case is plotted again in Fig. \ref{EPDs}(a), but tracking each individual solution of the dispersion equation, including where the longitudinal modal wavenumber becomes complex, as frequency is varied, and assuming this is a 1D propagation scenario where $k_y$ is forced to be zero. Mathematically, solutions of the dispersion equation cannot simply disappear as frequency (assumed real) is varied; therefore, above the inflexion point, solutions still exist but they are expected to correspond to evanescent modes. In other words, the inflexion point acts as a cut-off frequency or band edge and, as seen in Fig. \ref{EPDs}(a), the local dispersion around this point resembles the typical \emph{square-root-like bifurcation} around an exceptional point of degeneracy (EP) \cite{Capolino,Hanson_EPD}. 

To clarify this behavior, Fig. \ref{EPDs}(b,c,d) show the modal solutions (SPP poles) on the complex $k_x$-wavenumber plane as frequency is increased. At frequencies lower than the inflexion point, i.e., $ \omega < \omega_{inf} $, owing to the downward bending of the dispersion curve we have two counter-propagating modes (opposite group velocity) on the positive side of the real $k_x$ axis (Fig. \ref{EPDs}(b)). Right at the inflexion point, these two modes/poles coalesce into a single mode with vanishing group velocity (Fig. \ref{EPDs}(c)). Finally, for frequencies higher than the inflexion-point, $ \omega > \omega_{inf} $, the poles rapidly migrate away from the real axis (Fig. \ref{EPDs}(d)), leading to an interesting modal transition further discussed below. This behavior indeed suggests that the inflexion point could also be interpreted, in this 1D propagation scenario, as an exceptional point of degeneracy since it marks the coalescence and branching of modes \cite{Capolino,Hanson_EPD}. 

To further confirm this interpretation, we considered the general conditions that a solution of the dispersion equation, in a 1D wave-guiding system, must satisfy to be identified as an EP, as originally derived in \cite{RS_1998,RS_1999,Hanson_TAP_2003}. The first necessary condition to obtain a first-order EP is that two first-order roots of the dispersion equation, $D(k_{\mathrm{EP}}, \omega_{\mathrm{EP}}) = 0$, coalesce to form a second-order root, which implies that, at the EP,
\begin{equation}\label{1st_cindition}
	{\left. {D(k_{\mathrm{EP}}, \omega_{\mathrm{EP}}) = \frac{\partial D(k, \omega)}{ \partial k}} \right|_{\scriptstyle k = {k_{\mathrm{EP}}}\hfill\atop \scriptstyle\omega  = {\omega _{\mathrm{EP}}}\hfill}} = 0
\end{equation}
where $ k_{\mathrm{EP}}$ and $  \omega_{\mathrm{EP}}  $ are the wavenumber and angular frequency at which the degeneracy emerges. However, an EP should also satisfy an additional condition,
\begin{equation}\label{2nd_cindition}
	{\left. {\frac{\partial D(k, \omega)}{ \partial \omega} \cdot \frac{\partial^2 D(k, \omega)}{ \partial k^2}} \right|_{\scriptstyle k = {k_{\mathrm{EP}}}\hfill\atop \scriptstyle\omega  = {\omega _{\mathrm{EP}}}\hfill}} \neq 0.
\end{equation}
In fact, only if Eq. (\ref{2nd_cindition}) is satisfied, then the second-order root becomes a branch point (and not a saddle point) where various branches of the dispersion function $k(\omega)$ merge \cite{RS_1998,RS_1999,Hanson_TAP_2003}. Our numerical tests applied to the case of SPPs propagating along the electron current verify that the local 1D dispersion function, $k_x(\omega)$ (for $k_y=0$), satisfies conditions (\ref{1st_cindition}) and (\ref{2nd_cindition}) around the inflexion frequency (red point in Fig. \ref{Fig2}(a)), hence confirming that in this 1D scenario, the inflexion point can be interpreted as an EP (branch point) where forward-propagating and backward-dragged SPPs coalesce. We also note that, unlike typical EPs in open non-Hermitian systems, the EP observed here exists in a system with no balanced distribution of loss and gain. Indeed, this dispersion feature is more similar to the EPs associated with the band edges of lossless periodic structures (see, for example, Refs. \cite{Figotin, Miri,Hanson_EPD}); however, in the considered current-biased system, the EP exists at a sort of unidirectional band edge (the inflexion point) created not by a periodic modulation, but by the Doppler shift resulting from the applied current.

The modal transition occurring through this EP is particularly interesting as it leads to a rich and subtle behavior at higher frequencies, and some additional observations should be made. As discussed in Ref. \cite{FM_Nature}, in a system with continuous translational symmetry as the one considered here, a \emph{direct} transition from a purely propagating solution to a purely evanescent one, or viceversa, can only occur at band-edges or cut-off frequencies at zero or infinite wavenumbers (where the modal wavenumber can go \emph{directly} from purely real to purely imaginary). This is clearly not the case in Fig. \ref{EPDs}(a), where the band-edge (exceptional point) exists in the dispersion diagram at a finite non-zero value of wavenumber and the solutions above the EP frequency are indeed complex, with the same real part but opposite imaginary parts. However, since we assumed that the considered structure is lossless, $\Gamma=0$, a complex wavenumber is puzzling because it implies some form of energy dissipation or gain as the wave propagates. 
One of the two complex solution branches above the EP in Fig. \ref{EPDs}(a) represents an exponentially growing propagating wave, which, we argue, is a non-physical solution since the considered system does not provide optical gain. Yet, due to the presence of the drift current, this point deserves further clarification.  
Indeed, a direct current bias may in principle lead to optical gain through negative Landau damping in an otherwise passive system:  if the drift velocity is comparable and greater than the wave phase velocity, $ v_d > \omega / k $, the electrons ``trapped'' inside the moving potential wells of the wave lose kinetic energy to the wave (the opposite process is the regular Landau damping discussed in the following) \cite{Bittencourt,Mario-Active}. However, this cannot happen in the present situation since the SPP phase velocity in the region of interest of the dispersion diagram is much larger than $v_d$. In addition, it was convincingly demonstrated in Ref. \cite{Mario-Negative-Landau} that SPP amplification and instabilities are not possible in a configuration with a single drift-biased plasmonic medium. This also makes sense physically by considering the analogy with moving media: a single body moving at a constant velocity cannot lead to wave instabilities; instead, the presence of another body in close proximity and in relative motion is necessary, for example two graphene sheets with different bias velocities \cite{Mario-Negative-Landau}, or a single drift-biased graphene sheet on a non-biased SiC substrate \cite{Mario-Active}. Based on these considerations, we can safely say that the case considered in this section, with a single drift-biased material, cannot lead to amplification and instabilities, and any unstable pole can be considered unphysical. Indeed, the presence of complex poles with opposite imaginary parts is common in even simpler passive systems, for example beyond a band-edge in the gapped band diagram of a periodic structure, or below the cut-off frequency of a waveguide. Mathematically, poles on both sides of the real axis do exist in these scenarios, but only the exponentially decaying solution is retained on physical grounds (i.e., the proper branch of a multivalued dispersion relation is chosen based on considerations of physical realizability).

Going back to our discussion of the dispersion diagram in Fig. \ref{EPDs}(a), we note that the other solution branch above the EP corresponds, instead, to a physical, decaying, propagating wave that becomes overdamped as it crosses the light line at higher frequencies, which suggests that radiation leakage plays an important role in this behavior. In fact, analogous modal transitions (albeit for reciprocal systems) were observed and investigated in the context of leaky-wave antennas a few decades ago \cite{Oliner1,Oliner2}. In the transition region between bound- and leaky-waves, it was similarly observed that certain propagating solutions are non-physical (growing in the longitudinal direction), and the unexpectedly lossy solutions below the light line, in an otherwise lossless structure, can be explained as the ``winding down'' of the leaky-wave solution right under the light cone (with a virtually negligible contribution to the total fields excited by an actual source \cite{Oliner2}).


\begin{figure}[h!]
	\centering\includegraphics[width=\columnwidth]{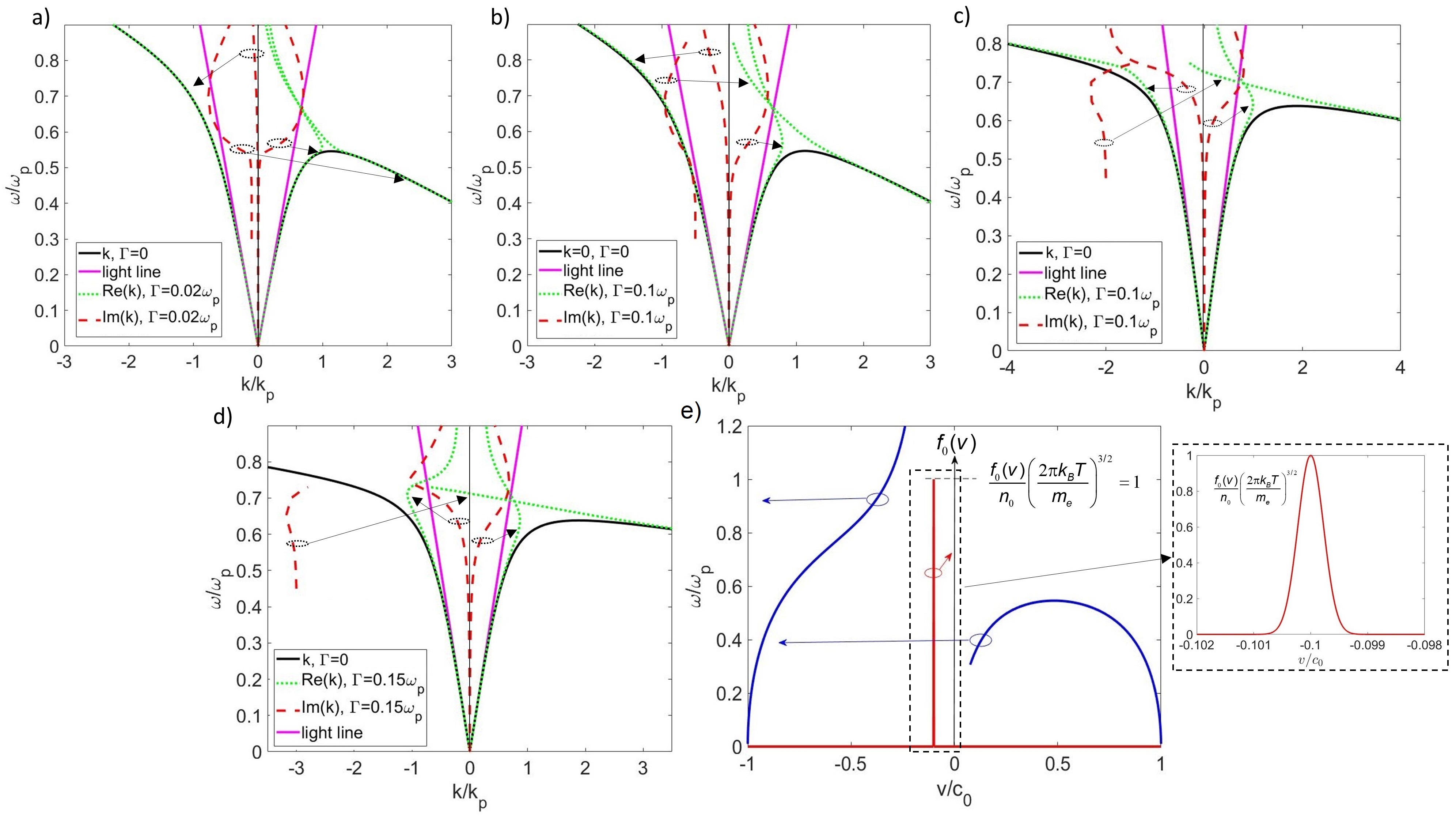}
	\caption{Impact of dissipation on nonreciprocal current-biased SPPs. (a,b) Dispersion curves for SPPs propagating along the current flow, with $ v_d=-c/10 $ in the presence of intrinsic (bulk) scattering losses: (a) $ \Gamma=0.02\omega_p $ and (b) $ \Gamma = 0.1\omega_p $. (c,d) Dispersion curves for SPPs propagating along the current flow, with $ v_d=-c/40 $ and (c) $ \Gamma=0.1\omega_p $ and (d) $ \Gamma = 0.15\omega_p $. 
		For the forward-propagating mode with largest wavevector, the imaginary part is plotted over a limited frequency range due to numerical difficulties in tracking this solution beyond this range. Solid black lines denote the lossless case (showing only the real branches for simplicity), and dotted green and dashed red lines indicate the real and imaginary part of the in-plane modal wavenumber, $ \mathrm{Re}(k) $ and $ \mathrm{Im}(k) $, respectively. Arrows indicate which imaginary and real branches correspond to the same mode. Solid purple lines denote the light cone. (d) Phase velocity of SPP modes propagating parallel to the drifting electrons with $ v_d = -c/10 $, compared with a Maxwell-Boltzman velocity distribution for a typical plasmonic material at room temperature and with the same electron drift velocity. The inset shows a zoomed-in view of the distribution function.
	}\label{lossy}
\end{figure}

Finally, we investigated how the dispersion diagram changes when absorption losses are included in the current-biased plasmonic material. This is particularly important considering the virtually unavoidable presence of dissipation mechanisms in plasmonic media. In particular, here we first studied the impact of the intrinsic (bulk) scattering losses of the material, modeled by an intrinsic damping rate $\Gamma$ which is assumed to be unaltered by the presence of the current bias, as done for example in \cite{Mario-Active}. However, how this damping rate affects the propagation and attenuation of SPPs strongly depends on the presence and velocity of the drift current, the SPP wavevector, and the specific structure. 
%
%
As seen in Fig. \ref{lossy}(a,b), increasing $\Gamma$ determines an increase in the imaginary part of the modal wavenumber, especially in the large-wavevector regime, as expected. Non-zero dissipation also ``opens'' the band bifurcation at the EP, such that the dispersion curves no longer flatten out completely. This behavior is therefore qualitatively different from the way losses affect nonreciprocal SPPs in magnetically biased plasmas \cite{FM_Nature, Hassani_optica}, where band edges are located at diverging values of wavenumber. As a result of this modified dispersion diagram, the effects described in Section 2.1, which mostly depend on a slowing down of the group velocity at the inflexion point, are expected to be sensitive to the presence of dissipation (which may actually be useful for the sensing of absorbing layers on the plasmonic platform). Figure \ref{lossy}(b,c) compare two cases where $ \Gamma $ is kept fixed but the drift velocity is decreased. As seen here, by reducing the drift velocity, the imaginary part of the modal wavenumber generally increases, indicating that, in this configuration, drift-induced nonlocality and nonreciprocity may indirectly mitigate the detrimental impact of scattering losses by modifying the dispersion diagram and the group velocity of the modes. 
Finally, if losses are sufficiently large and the drift velocity is low (Fig. \ref{lossy}(d)), dissipation starts to dominate over the drift-induced nonlocal and nonreciprocal effects. In particular, the backward-propagating mode now exhibits the typical ``back-bending'' of a lossy SPP (real dispersion curve bends in the opposite direction) and becomes overdamped at higher frequencies ($\text{Re}[k]<\text{Im}[k]$), whereas it was still underdamped in Fig. \ref{lossy}(c). The smaller-wavevector branch of the forward-propagating mode shows a similar back-bending behavior (as it does in all cases in Fig. \ref{lossy}), whereas the other, flatter branch becomes overdamped over a broad range of wavevectors (and for larger wavevectors it is essentially a dark mode, increasingly difficult to excite). Unidirectional propagation is no longer possible in this regime as the forward- and backward-propagating modes are both allowed to propagate at low frequencies and, as frequency increases, they become overdamped almost simultaneously. 

In addition to intrinsic scattering loss, another potentially important attenuation mechanism in drift-biased plasmas and plasmonic materials is bulk Landau damping, which exists even in the absence of any collisions, originating instead from the kinetic energy transfer from an electromagnetic wave to the electrons ``trapped'' inside its moving potential wells. This effect is possible if the velocity distribution function is not a delta function centered at the drift-bias velocity (cold biased plasma), but follows for example a Maxwell-Boltzman distribution $f_0(v)$ for a plasma at non-zero temperature. The attenuation due to bulk Landau damping can be shown to be proportional to the velocity derivative $ \partial_v f_0(v) $ \cite{Lifshitz,Bittencourt}, which indicates that this form of damping is important if the electron velocities are close to, and smaller, than the wave phase velocity. For the considered drift-biased nonreciprocal configuration, Figure \ref{lossy}(d) compares the phase velocity of SPPs propagating along the current with the Maxwell-Boltzman distribution function for a typical plasmonic material at 300K. With the exception of the backward-propagating mode at very large frequencies, the phase velocity of the SPP modes lies far out on the distribution tail, where $ \partial_v f_0(v) \simeq 0 $, which implies that the presence of bulk Landau damping can be safely ignored in the present case. We also note that another form of Landau damping, namely, surface-collision-induced damping, associated with the direct excitation of electron-hole pairs by the highly confined electric field on the interface \cite{PRL_Hassani_2}, may be more significant in this case, but is still expected to be smaller than the intrinsic (bulk) scattering losses of typical solid-state plasmonic media. 


The analysis presented in this section shows, for the first time to the best of our knowledge, the impact of dissipation (due to collisions or Landau damping) on the drift-induced nonlocal and nonreciprocal effects in current-biased 3D plasmonic platforms. While the presence of losses is detrimental for wave-guiding applications, the next section discuss an example of application of nonreciprocal drift-biased plasmonics where losses are actually necessary, namely, radiative heat transfer mediated by SPPs.

\section{Drift-Induced Nonreciprocal Radiative Heat Transfer}

\begin{figure*}[h!]
	\begin{center}
		\noindent
		\includegraphics[width=\columnwidth]{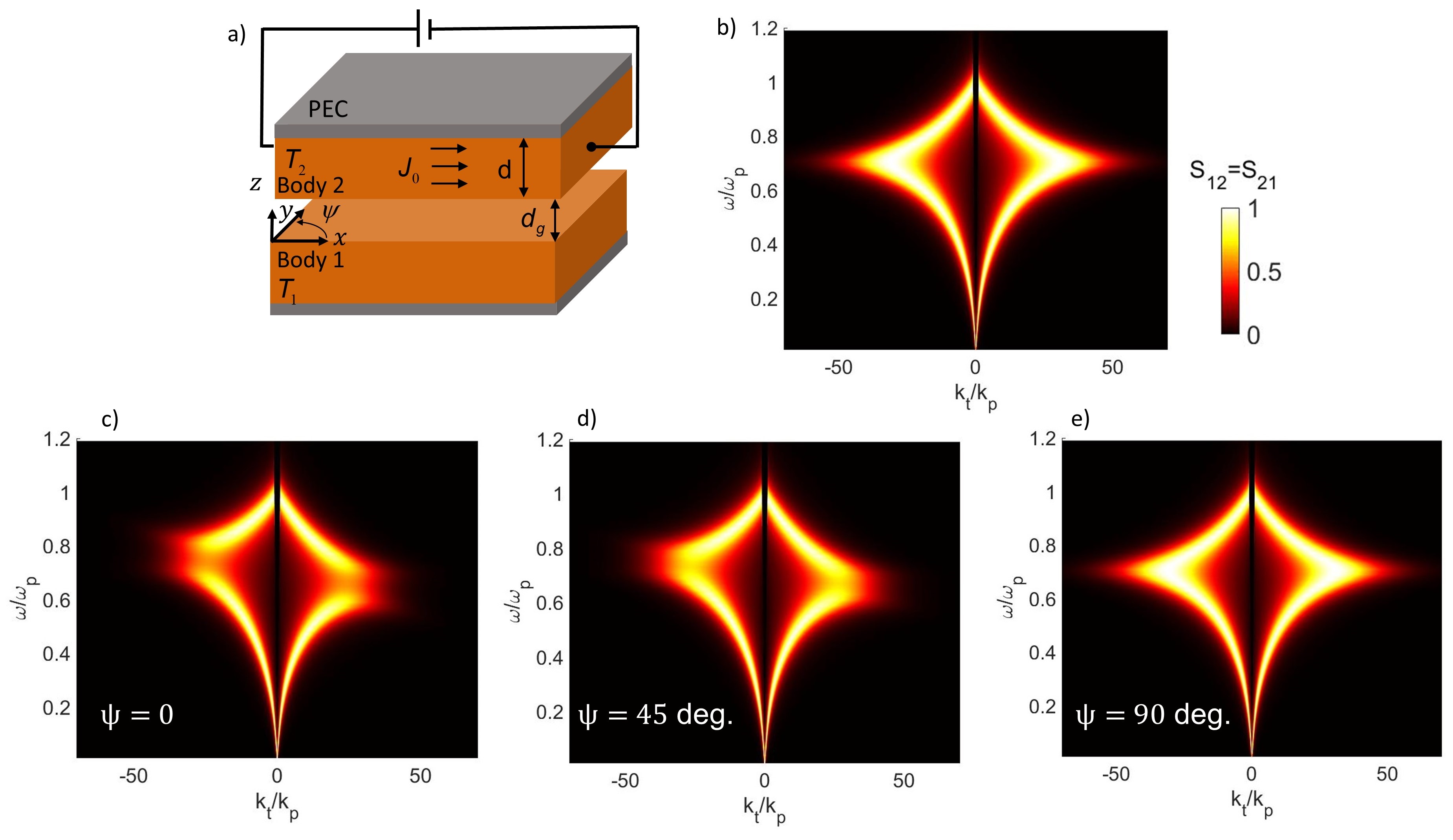}
		\caption{ (a) Illustration of the geometry considered for investigating the near-field radiative heat transfer between two planar bodies in the reciprocal and nonreciprocal (current-biased) cases. Two planar conducting slabs are placed at a distance $ d_g = \lambda_p/100 $ from each other and one of them is biased by a DC drift current. The thickness of the slabs is $ d = \lambda_p $, where $ \lambda_p $ is the free-space wavelength at the plasma frequency. The two slabs are supposed to have identical plasma frequency and damping rate, $ \Gamma = 0.1 \omega_p $. (b) Heat flux density between the two bodies as a function of frequency and in-plane wavenumber when the current bias is set to zero. Identical heat flux density maps are obtained in this case for any in-plane direction defined by the angle $\Psi$ with respect to the $+x$-axis. (c,d,e) Heat flux density for the same configuration but with the top conducting body biased by a DC current with $ v_d = -c/300 $ along the $+x$-axis. The three panels show the heat flux density map for different in-plane directions. While the system being studied is, strictly speaking, an active non-equilibrium system, we operate in a regime with relatively large separations and low velocities, where non-equilibrium phenomena are still small, but nonreciprocal effects are clearly visible, as further discussed in the text.  }\label{Fig5}
	\end{center}
\end{figure*}

We consider the canonical problem of near-field radiative heat transfer between two planar bodies, as shown in Fig. \ref{Fig5}(a): two plasmonic slabs, with permittivity $ \epsilon(\omega) = 1 - (\omega_{p}/\omega)^2/(1+i\Gamma/\omega)$, are separated by a distance $d_g$, and backed by perfect electric conductors. A DC electric current is driven in the top conductor to break reciprocity. As discussed in the previous section, the current-biased plasmonic material can then be approximately described by a spatially and frequency dispersive Doppler-shifted permittivity function $ \epsilon(\omega) \rightarrow \epsilon(\omega - \boldsymbol{k} \cdot \boldsymbol{v}_d) $. 
The geometry being studied and the following calculations parallel the ones in Ref. \cite{Fan-S12}, but with the crucial difference that here we consider, for the first time, a direct current bias to realize nonreciprocal radiative heat transfer, instead of magnetized magneto-optical materials. 

Before discussing our results in details, an important point should be clarified: in general, the systems being studied here are non-equilibrium systems due the presence of drifting electrons. Indeed, considering the analogy with moving media, even in the limit of zero temperature two bodies in relative motion and close proximity can exchange energy, which can be explained in terms of quantum friction \cite{Pendry,Mario-quantum-friction}, also related to the possible onset of wave instabilities in moving media (and negative Landau damping in drift-biased plasmas) mentioned in the previous section. Moreover, strictly speaking, the fluctuation-dissipation theorem \cite{Landau-2} applies to systems at equilibrium. Fully accounting for the non-equilibrium thermodynamics of such systems is quite involved, as discussed for example in Ref. \cite{Nonequilibrium}. Importantly, however, these effects become significant only when the media/sources in relative motion are in very close proximity (usually atomic distances as in \cite{Pendry}) and/or for very large relative velocities (not necessarily relativistic). More precisely, as shown in Ref. \cite{Nonequilibrium}, large deviations from the equilibrium fluctuation-dissipation theorem occur for $\omega \ll v/d_g$ (i.e., $2\pi d_g/\lambda \ll v/c$), where $d_g$ is the separation between the moving bodies. Here, instead, we focus on a different regime, with relatively large separations and low velocities, where non-equilibrium effects are still small, but nonreciprocal properties are already clearly visible.

Under the assumptions discussed above, the equilibrium fluctuation-dissipation theorem \cite{Landau-2} implies that the strength of the fluctuating current sources that generate thermal radiation is proportional to the imaginary part of the permittivity, $ \mathrm{Im}\left( \epsilon \right) \equiv \left(  \epsilon - \epsilon^* \right)/2i $. Unlike many areas of photonics in which reducing material absorption is an important goal, thermal photonic applications do require materials with non-zero losses since a lossless structure does not generate thermal emission. This relation between absorption and emission is however a subtle one in nonreciprocal media, as discussed in the following. In the setup in Fig. \ref{Fig5}(a), we are interested in the radiative heat flux density from body $i$ to body $j$, $S_{ij}$, and vice versa, which is defined as the power density absorbed by body $j$ from the radiated electric and magnetic fields generated by the fluctuating currents in body $i$. The total heat flux density from body 1 to body 2 can be expanded in terms of its time-harmonic plane-wave contributions, i.e., a continuous set of ``channels'' defined by their wavevector, $\boldsymbol{k} = \pm(\boldsymbol{k}_z+\boldsymbol{k}_t$), and frequency, $\omega$, \cite{Fan-S12}
\begin{equation}\label{flux_density}
	S_{12} = \int_{0}^{\infty} \frac{d \omega}{2\pi} \int \frac{d \boldsymbol{k}_t}{\left(2\pi\right)^2}\Phi(T_1, \omega) S_{12}(\boldsymbol{k}_t, \omega)
\end{equation}
where $ \Phi(T, \omega) = \hbar \omega \left[ \frac{1}{2} + \frac{1}{ \mathrm{exp}\left( \hbar \omega / K_B T \right) - 1 }    \right] $ is the mean energy of photons per frequency $ \omega $ at temperature $T$ and 
\begin{align}\label{flux_density_1}
	S_{12}(\boldsymbol{k}_t, \omega) & = \mathrm{Tr} \left(\left[\hat{\boldsymbol{I}} - \hat{R}_2^{\dagger}(\boldsymbol{k}_t, \omega) \hat{R}_2(\boldsymbol{k}_t, \omega)\right] \hat{D}_{12}(\boldsymbol{k}_t, \omega) \right. \nonumber \\& \left.  \times \left[\hat{\boldsymbol{I}} - \hat{R}_1(\boldsymbol{k}_t, \omega) \hat{R}_1^{\dagger}(\boldsymbol{k}_t, \omega)\right] \hat{D}_{12}^{\dagger}(\boldsymbol{k}_t, \omega)\right)
\end{align}
for propagating waves, $ \left| \boldsymbol{k}_t \right| < \omega/c $, and
\begin{align}\label{flux_density_2}
	S_{12}(\boldsymbol{k}_t, \omega) & = \mathrm{Tr}\left(\left[ \hat{R}_2^{\dagger}(\boldsymbol{k}_t, \omega) -  \hat{R}_2(\boldsymbol{k}_t, \omega)\right] \hat{D}_{12}(\boldsymbol{k}_t, \omega) \right. \nonumber \\& \left.  \times \left[ \hat{R}_1(\boldsymbol{k}_t, \omega) -  \hat{R}_1^{\dagger}(\boldsymbol{k}_t, \omega)\right] \hat{D}_{12}^{\dagger}(\boldsymbol{k}_t, \omega) e^{-2\alpha d_g}\right)
\end{align}
for evanescent waves, $ \left| \boldsymbol{k}_t \right| > \omega/c $. The heat flux density from body 2 to body 1 can be obtained from Eqs. (\ref{flux_density_1}) and (\ref{flux_density_2}) by exchanging the subscripts 1 and 2 and changing the temperature in Eq. (\ref{flux_density}) from $T_1$ to $ T_2 $. In the above equations, $ \alpha $ is the decay rate of evanescent waves along the $z$-axis, $ R_i(\boldsymbol{k}_t, \omega) $ is the reflectivity matrix of body $i$, whose elements are the reflection coefficients for light incident from vacuum, for both transverse-electric and transverse-magnetic polarizations, and the parameter $\hat{D}_{12} = \left[ \hat{\boldsymbol{I}} - \hat{R}_2^{\dagger}(\boldsymbol{k}_t, \omega) \hat{R}_1(\boldsymbol{k}_t, \omega) e^{-i2k_zd_g}  \right]$ represents multiple reflections between the two bodies (see Ref. \cite{Fan-S12} for more details). This formalism allows us to calculate the near-field 3D heat transfer between two planar slabs with and without a current bias. As in our 3D Green function formalism in Section 2 and Supplementary Note III, the presence of a non-zero drift current modifies the permittivity, making it wavevector-dependent, and therefore changes the reflection coefficients for plane-wave incidence.


A nonzero net heat flow in a two-body geometry as in Fig. \ref{Fig5}(a) requires a temperature gradient, as dictated by the Second Law of Thermodynamics. In other words, if the two bodies are at the same temperature, the total heat flux densities between them are identical, $ S_{12}=S_{21} $ (integrated over all frequencies and wavevectors), independent of whether the system is reciprocal or not. It is then possible to show that thermodynamics imposes the same symmetry also for the individual contributions, i.e., $S_{12}(\boldsymbol{k}_t, \omega) = S_{21}(\boldsymbol{k}_t, \omega) $, again independent of reciprocity or the lack thereof. Interestingly, Ref. \cite{Fan-S12} demonstrated that reciprocity imposes another symmetry constraint for opposite in-plane wavevectors: $S_{12}(\boldsymbol{k}_t, \omega) = S_{12}(-\boldsymbol{k}_t, \omega) = S_{21}(-\boldsymbol{k}_t, \omega)$. In a nonreciprocal system, this symmetry constraint can be violated since the dispersion diagram is no longer necessarily symmetric, $\omega(\boldsymbol{k}_t) \neq \omega(-\boldsymbol{k}_t) $. As a result, $S_{12}(\boldsymbol{k}_t, \omega) \neq S_{12}(-\boldsymbol{k}_t, \omega) = S_{21}(-\boldsymbol{k}_t, \omega)$, which implies that if heat transfer from body 1 to body 2 occurs through a certain wavevector-channel, $\boldsymbol{k}=\boldsymbol{k}_z+\boldsymbol{k}_t$, the transfer from body 2 to 1 does not have to occur symmetrically through the time-reversed channel, $\boldsymbol{k}_{\text{TR}}=-\boldsymbol{k}_z-\boldsymbol{k}_t=-\boldsymbol{k}$, whose contribution can even be zero; the reverse transfer can now occur entirely through a different channel with the same $\boldsymbol{k}_t$, i.e., the specular one, $\boldsymbol{k}_{\text{S}}=-\boldsymbol{k}_z+\boldsymbol{k}_t$. This results in a persistent heat current between the two bodies even when they are kept at the same temperature \cite{Fan-S12_PRL,Fan-S12}. In other words, while a nonreciprocal body may emit strongly in a certain direction defined by $ \boldsymbol{k}$ (at a certain frequency), it could be designed to absorb very weakly in that same direction (i.e., for $-\boldsymbol{k}$) and frequency, corresponding to a breaking of Kirchhoff's law of thermal radiation (equal absorptivity and emissivity at each direction and frequency) \cite{Fan-Joule, Kirchhoff, Plank,Siegel}. We stress again that the existence of a persistent heat current for bodies at the same temperature or a violation of Kirchhoff's law in these systems does not break thermodynamic laws, since the total net heat flow into each body is still zero when the bodies have the same temperature. In the following, we provide the first theoretical demonstration of these effects in a platform that does not require magneto-optical or space-time-modulated media.

First we consider the reciprocal scenario with no current bias in either the lower or upper body. In this situation the material permittivity takes a local and isotropic form, where we assume the same plasma frequency, $\omega_p$, and scattering rate, $ \Gamma = 0.1\omega_p $, for both materials. Figure \ref{Fig5}(b) shows the heat flux density at each frequency and in-plane wavenumber. As expected for reciprocal materials, the heat flux density is symmetric with respect to in-plane wavenumber, $S_{12}(\boldsymbol{k}_t, \omega) = S_{12}(-\boldsymbol{k}_t, \omega) $, and the same plot is obtained for any in-plane direction $\Psi$ since the considered material is homogeneous and isotropic. We also verified that this wavevector- and frequency-resolved heat transfer map from body 1 to 2, $S_{12}(\boldsymbol{k}_t, \omega)$, is identical to the corresponding one from body 2 to 1, $S_{21}(\boldsymbol{k}_t, \omega)$, therefore respecting the Second Law. We also note that the specific shape of the heat flux density map in Fig. \ref{Fig5}(b) is a result of coupling between the surface waves supported by the two interfaces, which strongly modifies their individual dispersion diagram. 
For subwavelength gaps between the two slabs, as in Fig. \ref{Fig5}, the structure supports strongly confined SPPs, for which even relatively small values of drift velocity determine large changes in their response, as seen in our results discussed in the following.

We make the system nonreciprocal by biasing the top conductor with a direct current with $ v_d = c/300 $ (much lower than in previous sections) along the $ +x $-axis. The heat transfer spectra for this nonreciprocal system are shown in Figs. \ref{Fig5}(c,d,e) for different in-plane wavevector directions, defined by the angle $\Psi$ with respect to the current flow. As seen from these results, $ S_{12}(\boldsymbol{k}_t, \omega) \neq S_{12}(-\boldsymbol{k}_t, \omega) $ for any in-plane direction except at $ \Psi = 90 $ deg., which is consistent with our previous observation that there is no interaction between surface waves and drifting electrons for wavevectors normal to the current flow. In this latter situation, the heat flux density map is exactly the same as in the reciprocal scenario in Fig. \ref{Fig5}(b). These results clearly demonstrate that such a magnet-free current-biased system exhibits a signature of nonreciprocity for near-field radiative heat transfer, which implies, as mentioned above, the presence of a persistent heat current between the two bodies even when they are kept at the same temperature. The asymmetric heat transfer spectra in Figs. \ref{Fig5}(c,d) are similar, but not exactly identical, to the ones obtained with magneto-optical materials in \cite{Fan-S12}. In particular, the heat transfer spectra in \cite{Fan-S12} follow the characteristic magnetic-field-induced asymmetry in the flat dispersion asymptotes of surface magneto-plasmons, whereas our results in Figs. \ref{Fig5}(c,d) show an asymmetry that arises from a Doppler-shift-induced tilt of the dispersion asymptotes. 
Furthermore, we have again verified that the current-biased system still respects the constraint $ S_{12}(\boldsymbol{k}_t, \omega) = S_{21}(\boldsymbol{k}_t, \omega)  $, which needs to be satisfied regardless of the local or nonlocal and reciprocal or nonreciprocal nature of the system.

We reiterate that the results and considerations above are valid as long as we operate in a regime where non-equilibrium effects due to the drifting electrons are negligible (the equilibrium fluctuation-dissipation theorem applies if $\omega \gg v/d_g~$ \cite{Nonequilibrium}). Indeed, the distance between the two slabs in Fig. \ref{Fig5} is much larger than the atomic distances considered in the study of quantum friction in Ref. \cite{Pendry} (if graphene was used as the plasmonic material, operating at mid-infrared wavelengths, the distance in Fig. \ref{Fig5} would be on the order of 100 nm, three orders of magnitude larger than in Ref. \cite{Pendry}). Moreover, while the two-body system being studied is active and could, in principle, exhibit optical gain, negative Landau damping, and instabilities, as demonstrated for similar geometries in Refs. \cite{Mario-Negative-Landau,Mario-Active} , we have verified that, in our case, these effects are negligible as the distance between layers is two orders of magnitude larger ($d_g \approx 10^{-2} \lambda $ for the case studied here, whereas $d_g \approx 10^{-4} \lambda $ in Ref. \cite{Mario-Negative-Landau}), for the same relative drift velocity. The analysis of small nonequilibrium corrections to our results and, more broadly, the full study of the nonequilibrium thermodynamics of such a nonreciprocal active system for smaller distances and higher velocities is potentially a very interesting area of research and will be the subject of future investigations.

\section{Conclusion}

The possibility of breaking reciprocity using a direct drift-current bias in conducting materials has been known for some time, but is starting to receive more attention only recently. The existing literature has been mostly focused on unidirectional propagation effects along the direction of the current and/or on specific platforms such as graphene. Here, instead, using a 3D Green function formalism and microscopic considerations (starting from the Vlasov equation, as further discussed in the Supplemental Material), we have studied current-induced nonreciprocal effects on the entire two-dimensional surface of a generic three-dimensional plasmonic platform, elucidating the dispersion and propagation properties of nonreciprocal SPPs and revealing a number of anomalous and extreme wave-propagation effects beyond unidirectional propagation, including the possible excitation of frozen-light, steerable, surface-wave beams with enhanced and localized fields, and the presence of exceptional-point-like modal transitions at inflexion points. Since plasmonic media are inherently associated with absorption losses, we have also clarified the impact of dissipation (due to collisions and Landau damping) on nonreciprocal current-biased SPPs and the associated tradeoffs. 

We have then focused on a particularly relevant example of application of these concepts, namely, the problem of breaking reciprocity for radiative heat transfer, which is usually achieved using magneto-optical materials. We have theoretically demonstrated a clear signature of nonreciprocity in this context, namely, an asymmetry in the radiative heat flux density between two planar bodies for opposite in-plane wavevectors, by relying, for the first time, on a direct drift-current bias instead of a magnetic bias. Our findings may open new opportunities and directions toward the long-sought goal of controlling radiative heat transfer without the constraints of time-reversal symmetry and reciprocity. 
Importantly, nonreciprocity in this context can be obtained with relatively low values of drift velocity, which could be practically realizable in certain high-mobility plasmonic media, such as high-mobility semiconductors and graphene in the THz, far-IR, and mid-IR spectral range. To demonstrate the nonreciprocal effects discussed in this paper, these are much more promising material platforms than noble metals, such as gold and silver, which have relatively low electron mobility. As an example, while in Ref. \cite{Bliokh} a gold nanowire carrying a realistic electric current on the order of tens of mA was predicted to support a modest drift velocity $v_d/c \approx {10}^{-6}$, producing very small nonreciprocal effects, if the same nanowire was made of high-mobility InSb, which exhibits plasmonic response in the low THz range \cite{InSb}, drift velocities up to $v_d/c\ \approx{10}^{-3}$ would be achievable, comparable to the values considered in this paper, for even smaller currents. Finally, thanks to its ultra-high electron mobility and mid-infrared plasmonic response, graphene is arguably the most interesting material for drift-biased nonreciprocal plasmonics\cite{Mario-Negative-Landau,Collimated-SPP,Basov}, enabling drift velocities on the order of its Fermi velocity, i.e., $v_d=c/300$.

In summary, we believe that the rich physics of drift-biased plasmonic systems, which enable nonreciprocal, slow-light, active, tunable effects at the nanoscale, may open up novel opportunities, combining the advantages of plasmonics and nonreciprocal electrodynamics, for both wave-guiding applications and thermal photonics.

\section*{Funding}

The authors acknowledge support from the Air Force Office of Scientific Research with Grant No. FA9550-19-1-0043 through Dr. Arje Nachman (whom we also thank for his suggestions on this manuscript) and the National Science Foundation with Grant No. 1741694.

\end{document}